\documentclass[twocolumn,tighten,twocolappendix]{aastex7}
\pdfoutput=1 
\usepackage{amsmath,amstext}
\usepackage[T1]{fontenc}
\usepackage{ae,aecompl}
\usepackage[utf8]{inputenc}
\usepackage{apjfonts} 
\usepackage[figure,figure*]{hypcap}
\usepackage{enumitem}
\usepackage{bm}
\usepackage{graphicx}
\usepackage{lineno}
\usepackage{tikz}

\input{hyperlink-year-only-natbib-patch}

\newcommand*{\http}[1]{\href{http://#1}{#1}}
\newcommand*{\https}[1]{\href{https://#1}{#1}}

\shorttitle{Enhanced $P(k)$ Substructure}
\shortauthors{Nadler, Gluscevic, \& Benson}

\begin{document}

\title{The Effects of Linear Matter Power Spectrum Enhancement on Dark Matter Substructure}

\author[0000-0002-1182-3825]{Ethan O.~Nadler}
\email{enadler@ucsd.edu}
\affiliation{Department of Astronomy \& Astrophysics, University of California, San Diego, La Jolla, CA 92093, USA}

\author[0000-0002-3589-8637]{Vera Gluscevic}
\email{vera.gluscevic@usc.edu}
\affiliation{Department of Physics $\&$ Astronomy, University of Southern California, Los Angeles, CA 90007, USA}

\author[0000-0001-5501-6008]{Andrew Benson}
\email{abenson@carnegiescience.edu}
\affiliation{Carnegie Observatories, 813 Santa Barbara Street, Pasadena, CA 91101, USA}

\correspondingauthor{Ethan~O.~Nadler}
\email{enadler@ucsd.edu}

\label{firstpage}

\begin{abstract}
We present cosmological dark matter (DM)--only zoom-in simulations of a Milky Way analog originating from enhanced linear matter power spectra $P(k)$ relative to the standard cold, collisionless DM (CDM) cosmology. We consider a Gaussian power excess in $P(k)$ followed by a cutoff in select cases; this behavior could arise from early-Universe physics that alters the primordial matter power spectrum or DM physics in the radiation-dominated epoch. We find that enhanced initial conditions (ICs) lead to qualitative differences in substructure relative to CDM. In particular, the subhalo mass function (SHMF) resulting from ICs with both an enhancement and cutoff is amplified at high masses and suppressed at low masses, indicating that DM substructure is sensitive to features in $P(k)$. Critically, the amplitude and shape of the SHMF enhancement depend on the wavenumber of the $P(k)$ excess and the presence or absence of a cutoff on smaller scales. These alterations to the SHMF are mainly imprinted at infall rather than during tidal evolution. Additionally, subhalos are found systematically closer to the host center, and their concentrations are increased in scenarios with $P(k)$ enhancement. Our work thus reveals effects that must be captured to enable $P(k)$ reconstruction using DM substructure.
\end{abstract}

\keywords{\href{http://astrothesaurus.org/uat/353}{Dark matter (353)}; 
\href{http://astrothesaurus.org/uat/1083}{$N$-body simulations (1083)};
\href{http://astrothesaurus.org/uat/1880}{Galaxy dark matter halos (1880)}}

\section{Introduction}\label{sec:intro}

Dark matter (DM) substructure is a powerful cosmological probe. The initial conditions (ICs) that seed DM subhalo populations are captured by the linear matter power spectrum $P(k)$, where $k$ is the wavenumber. Deviations in $P(k)$ from standard cold, collisionless DM (CDM) predictions on small scales can affect the mass function, density profiles, and redshift evolution of DM subhalo populations and the satellite galaxies they host (see \citealt{Bechtol220307354} for a review).

A wide range of physics can affect $P(k)$ on small scales. In particular, $P(k)$ depends on both the primordial matter power spectrum and its growth until recombination, captured by linear transfer functions. The primordial power spectrum is determined by the physics of inflation, while transfer functions capture the early expansion history and DM physics in the radiation-dominated epoch~(for reviews, see \citealt{Allahverdi200616182,Arbey210411488,Achucarro220308128}).

For example, many models beyond the CDM paradigm lead to a cutoff in $P(k)$ on sub-Mpc scales, which reduces the abundance of subhalos and satellite galaxies in the Milky Way (MW; \citealt{Nadler241003635}). On the other hand, the effects of $P(k)$ enhancement on subhalo populations are less well studied. More generally, understanding the mapping between features in $P(k)$ and DM subhalo populations is critical to enable tests of new physics using small-scale structure.

Structure formation in cosmologies with $P(k)$ enhancement is difficult to simulate once the relevant modes become nonlinear. In particular, unlike scenarios with a $P(k)$ cutoff, the growth of small-scale density perturbations is more difficult to capture accurately if there is substantial power on scales below the resolution limit of the simulation. Note that the dimensionless CDM power spectrum $k^3 P(k)$ grows logarithmically with increasing $k$; even for such mild growth toward small scales, unresolved small-scale power can affect the properties of structure on larger scales. A classic effect of unresolved small-scale power is that the innermost regions of simulated CDM halos have constant-density cores due to limited resolution~\citep{Power0201544}. These artificial cores become smaller as resolution is increased and give way to a steep power law ``prompt cusp'' if the first generation of halo collapse is captured~\citep{Delos220705082}. Because numerical disruption is accelerated in the presence of density profile cores~\citep{VandenBosch171105276,Errani200107077,Benson220601842}, unresolved small-scale power can thus impact the properties of subhalos that are nominally resolved.

Here, we study how subhalo populations respond to $P(k)$ enhancement using high-resolution simulations that accurately capture small-scale structure formation on scales relevant for MW satellites. As a benchmark example, we model $P(k)$ enhancement using a Gaussian excess (or ``bump'') that reaches $\approx 5\times$ the power in CDM at a chosen scale. Bumps in the primordial power spectrum are a hallmark of many inflationary models (e.g., with multiple fields or features in the inflaton potential; \citealt{Adams0102236,Palma200406106}) and can also arise from an early matter-dominated epoch, vector or axion DM, and primordial magnetic fields (e.g., \citealt{Erickcek11060536,Graham150402102,Buschmann190600967,Gorghetto200704990,Pavicevic250106299}). 

Power spectra with small-scale enhancement have garnered interest because they increase halo and galaxy abundances and inner densities at high redshifts~\citep{Liu220813178,Parashari230500999,Colazo240413110,Hirano230611993}. In particular, \cite{Tkachev240702991} and \cite{Eroshenko240902739} demonstrated that similar effects hold for the bumpy power spectra that we focus on here. As a result, \cite{Tkachev230713774} and \cite{Pilipenko240417803} showed that bumpy models are potentially in better agreement with high-$z$ JWST galaxy surveys~\citep{Finkelstein221105792,Labbe220712446} compared to standard CDM.

Building on the COZMIC simulations~\citep{An241103431,Nadler241003635,Nadler241213065}, we therefore run new zoom-in simulations of a MW analog to quantify how $P(k)$ bumps impact subhalo populations. In select cases, we also include cutoffs on scales smaller than the enhanced modes in $P(k)$ to mimic the effects of modeling these ICs with artificially truncated small-scale power. Specifically, we run nine new high-resolution cosmological DM--only zoom-in simulations of an MW analog originally from the Milky Way-est suite \citep{Buch240408043} with $P(k)$ bumps and/or suppression. This host was simulated with $P(k)$ suppression appropriate for warm, fuzzy, interacting, mixed, and self-interacting DM in the COZMIC suite~\citep{An241103431,Nadler241003635,Nadler241213065}.

Our results provide a benchmark for constraining $P(k)$ bumps using DM substructure. More generally, understanding the interplay between $P(k)$ and subhalo populations will be necessary to accurately interpret upcoming dwarf galaxy, stellar stream, and strong-lensing data in the context of early-Universe and DM physics. Indeed, recent simulations demonstrate that $P(k)$ modifications also affect substructure in cosmologies with a running spectral index (\citealt{Garrison-Kimmel14053985}, although these authors only considered a negative running), a blue-tilted primordial power spectrum (\citealt{Wu241216072}), or a population of primordial black holes (PBHs; \citealt{Colazo250523896}). To date, constraints on scalar spectral index running~\citep{Gilman211203293}, blue-tilted primordial power spectra ~\citep{Esteban230604674,Dekker240704198}, and PBHs~\citep{Balaji240811098} have relied on semianalytic predictions to connect $P(k)$ to subhalo populations. Calibrating such models on cosmological simulations is therefore timely.

This paper is organized as follows. In Section~\ref{sec:overview}, we describe our simulations. In Section~\ref{sec:results}, we study the subhalo mass function (SHMF); in Section~\ref{sec:drivers}, we dissect these results by studying isolated versus subhalo mass functions, subhalo radial distributions, and the subhalo mass--concentration relation. We summarize and discuss our results in Section~\ref{sec:discussion}.

We use cosmological parameters $h = 0.7$, $\Omega_{\rm m} = 0.286$, $\Omega_{\rm b} = 0.049$, $\Omega_{\Lambda} = 0.714$, $\sigma_8 = 0.82$, and $n_s=0.96$ (\citealt{Hinshaw_2013}). Virial masses are defined via $\Delta_{\mathrm{vir}}\approx 99\times \rho_{\mathrm{crit}}$, where $\rho_{\mathrm{crit}}$ is the critical density of the Universe at $z=0$~\citep{Bryan_1998}. We refer to halos within the virial radius of our MW host as ``subhalos,'' and halos outside the virial radius of any larger halo as ``isolated halos.''


\section{Simulation Overview}
\label{sec:overview}

\subsection{Initial Conditions}

We parameterize $P(k)$ for the Gaussian enhancement model as follows \citep{Tkachev240702991}
\begin{equation}
    \mathcal{T}_{\mathrm{bump}}(k) \equiv \sqrt{\frac{P_{\mathrm{bump}}(k)}{P_{\mathrm{CDM}}(k)}} = 1 + A\exp\left[-\frac{\left(\log(k)-\log(k_0)\right)^2}{\sigma_k^2}\right],
\end{equation}
where $\mathcal{T}(k)$ is the transfer function and $A$, $k_0$, and $\sigma_k$ are free parameters. Throughout, we set $A=2$ and $\sigma_k=0.5$; we describe our method for choosing $k_0$ below.

We parameterize $P(k)$ suppression using the thermal-relic WDM cutoff \citep{Vogel221010753}
\begin{equation}
    \mathcal{T}_{\mathrm{WDM}}(k) = \left[1+(\alpha(m_{\mathrm{WDM}})\times k)^{2\nu}\right]^{-5/\nu},\label{eq:transfer_wdm}
\end{equation}
where $m_{\mathrm{WDM}}$ is the WDM mass, $\nu=1.049$, and
\begin{equation}
\alpha(m_{\mathrm{WDM}})=a \left(\frac{m_{\mathrm{WDM}}}{1~\mathrm{keV}}\right)^b \left(\frac{\omega_{\mathrm{WDM}}}{0.12}\right)^{\eta}\left(\frac{h}{0.6736}\right)^{\theta}h^{-1}~\mathrm{Mpc},\label{eq:alpha_wdm}
\end{equation}
with $a=0.0437$, $b=-1.188$, $\theta=2.012$, and $\eta=0.2463$.

\begin{figure*}[t!]
\centering
\hspace{-5mm}
\includegraphics[trim={0 0.35cm 0 0cm},width=0.49\textwidth]{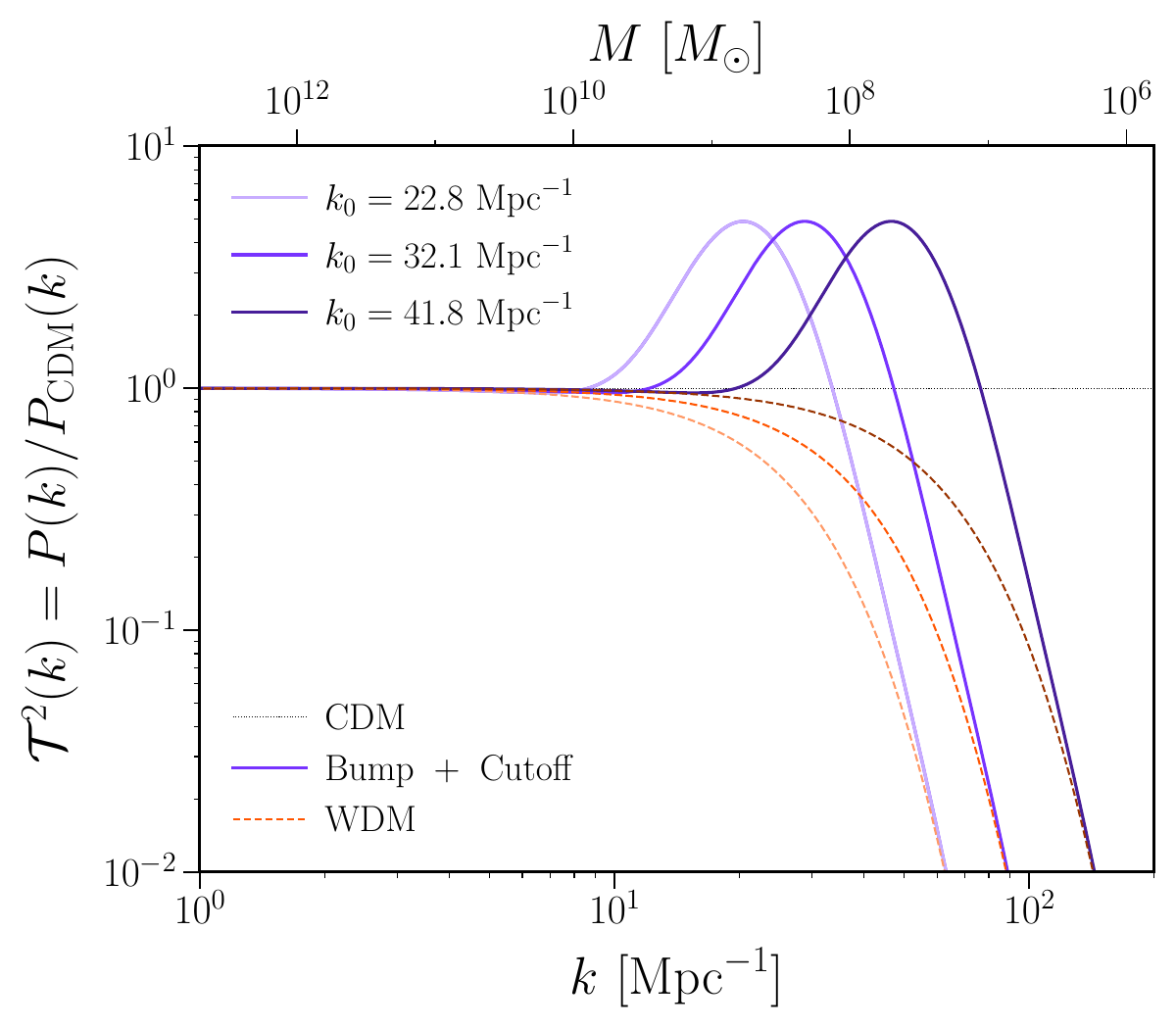}
\includegraphics[trim={0 0.35cm 0 0cm},width=0.49\textwidth]{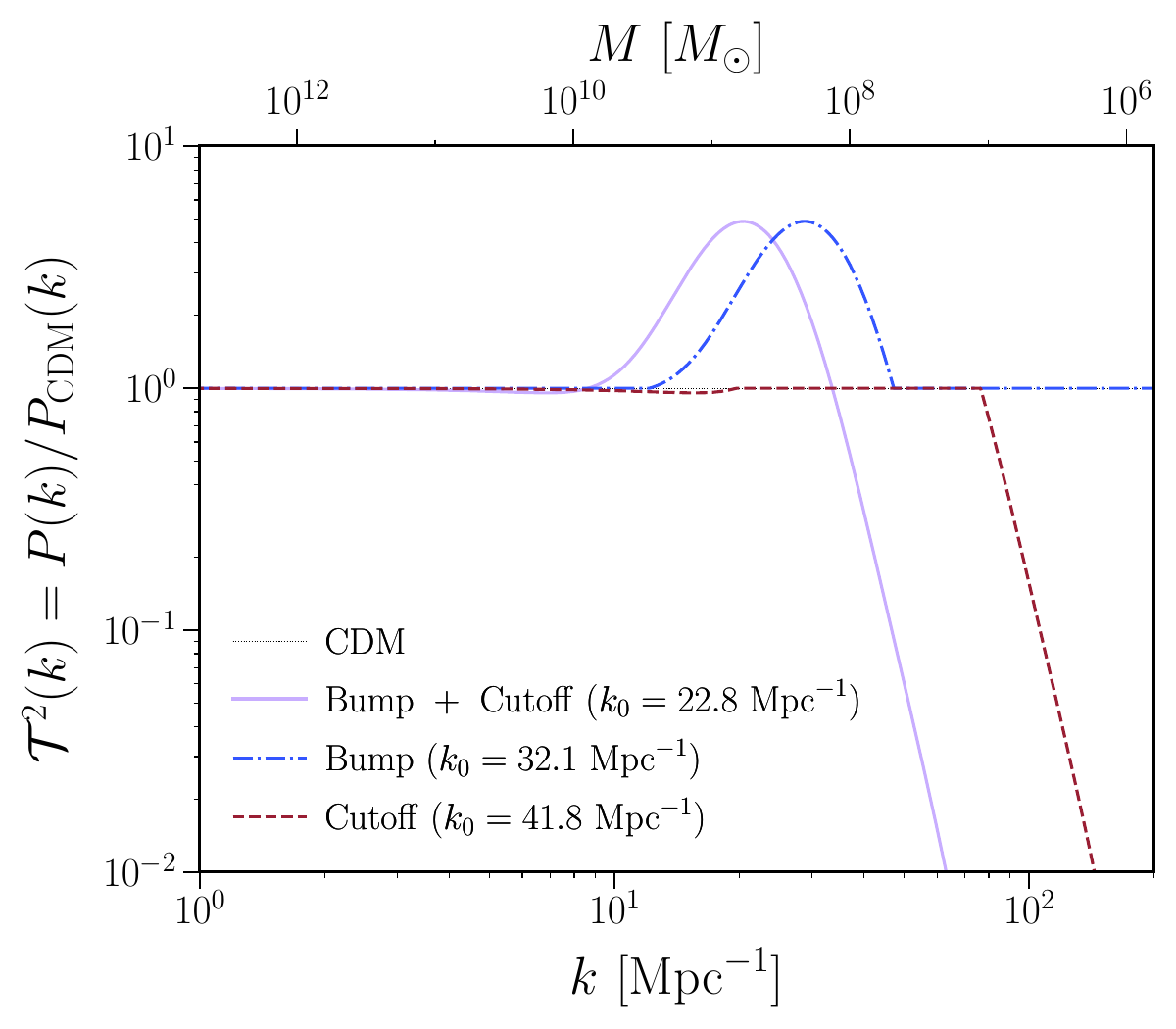}
    \caption{Left: ratio of the linear matter power spectrum in our Bump + Cutoff (solid purple) and WDM (dashed orange) scenarios relative to CDM (black dotted). Models are shown with $k_0=22.8$, $32.1$, and $41.8~\mathrm{Mpc}^{-1}$, corresponding to $k_{\mathrm{hm}}$ for $3$, $4$, and $5~\mathrm{keV}$ WDM, from lightest to darkest shade. Right: examples of Bump + Cutoff (solid purple), Bump (dotted--dashed blue), and Cutoff (dashed red) transfer functions with different values of $k_0$. In both panels, top ticks show halo masses associated with wavenumbers in linear theory~\citep{Nadler241003635}.}
\label{fig:transfer_enhanced}
\end{figure*}

We consider different combinations of $P(k)$ enhancement (``bumps'') and suppression (``cutoffs'') as follows:
\begin{equation}
\left\{
\begin{aligned}
&\mathcal{T}_{\mathrm{Bump+Cutoff}}(k) = \mathcal{T}_{\mathrm{Bump}}(k)\times \mathcal{T}_{\mathrm{WDM}}(k), && \text{Bump + Cutoff}; \\
&\mathcal{T}_{\mathrm{Cutoff}}(k) = \min\left[\mathcal{T}_{\mathrm{Bump+Cutoff}}(k), 1\right], && \text{Cutoff}; \\
&\mathcal{T}_{\mathrm{Bump}}(k)=\max\left[\mathcal{T}_{\mathrm{Bump+Cutoff}}(k), 1\right], && \text{Bump}.
\end{aligned}
\right.
\end{equation}
For the Bump + Cutoff, Bump, and Cutoff scenarios, we generate ICs for $k_0$ equal to the half-mode wavenumber of $3~\mathrm{keV}$ ($22.8~\mathrm{Mpc}^{-1}$), $4~\mathrm{keV}$ ($32.1~\mathrm{Mpc}^{-1}$), and $5~\mathrm{keV}$ ($41.8~\mathrm{Mpc}^{-1}$) WDM. Figure~\ref{fig:transfer_enhanced} shows the resulting ratios of $P(k)$ relative to CDM; the enhanced models reach about five times the power in CDM at their respective peaks, which occur at $\approx 0.9\,k_0$.

We feed the Bump + Cutoff, Bump, and Cutoff transfer functions into \textsc{MUSIC}~\citep{Hahn11036031} to generate zoom-in ICs. Our fiducial simulations use five refinement regions, resulting in a high-resolution particle mass of $5\times 10^4~M_{\mathrm{\odot}}$; we test for convergence in Appendix~\ref{sec:convergence}. We resimulate Halo004, and we use the COZMIC I CDM and WDM data presented at this resolution by \cite{Nadler241003635}. This host recently accretes an LMC analog and merges with a Gaia-Sausage-Enceladus analog at $z\approx 2$~\citep{Buch240408043}. The highest-resolution region spans 10 times the host's $263~\mathrm{kpc}$ virial radius at $z=0$.

\subsection{Zoom-in Simulations}

We run zoom-in simulations using \textsc{Gadget-2}~\citep{Springel0504097} with Plummer-equivalent gravitational softening of $80~\mathrm{pc}~h^{-1}$ in the highest-resolution region. Halo catalogs and merger trees are generated using \textsc{Rockstar} and \textsc{consistent-trees}~\citep{Behroozi11104372,Behroozi11104370}. We analyze only those subhalos above $300$ particles at $z=0$, corresponding to a virial mass threshold of $M_{\mathrm{vir}}>1.5\times 10^7~M_{\mathrm{\odot}}$; we define the peak mass $M_{\mathrm{peak}}\equiv \max(M_{\mathrm{vir}}(z))$. \cite{Nadler241003635} showed that the fraction of spurious subhalos formed through artificial fragmentation above this mass threshold is negligible for the $P(k)$ suppression scales we simulate, so we do not remove any (sub)halos from our halo catalogs after applying this $M_{\mathrm{vir}}$ cut.

\begin{figure*}[t!]
\centering
\vspace{0.5cm}
\includegraphics[trim={0 0cm 0 0cm},width=0.85\textwidth]{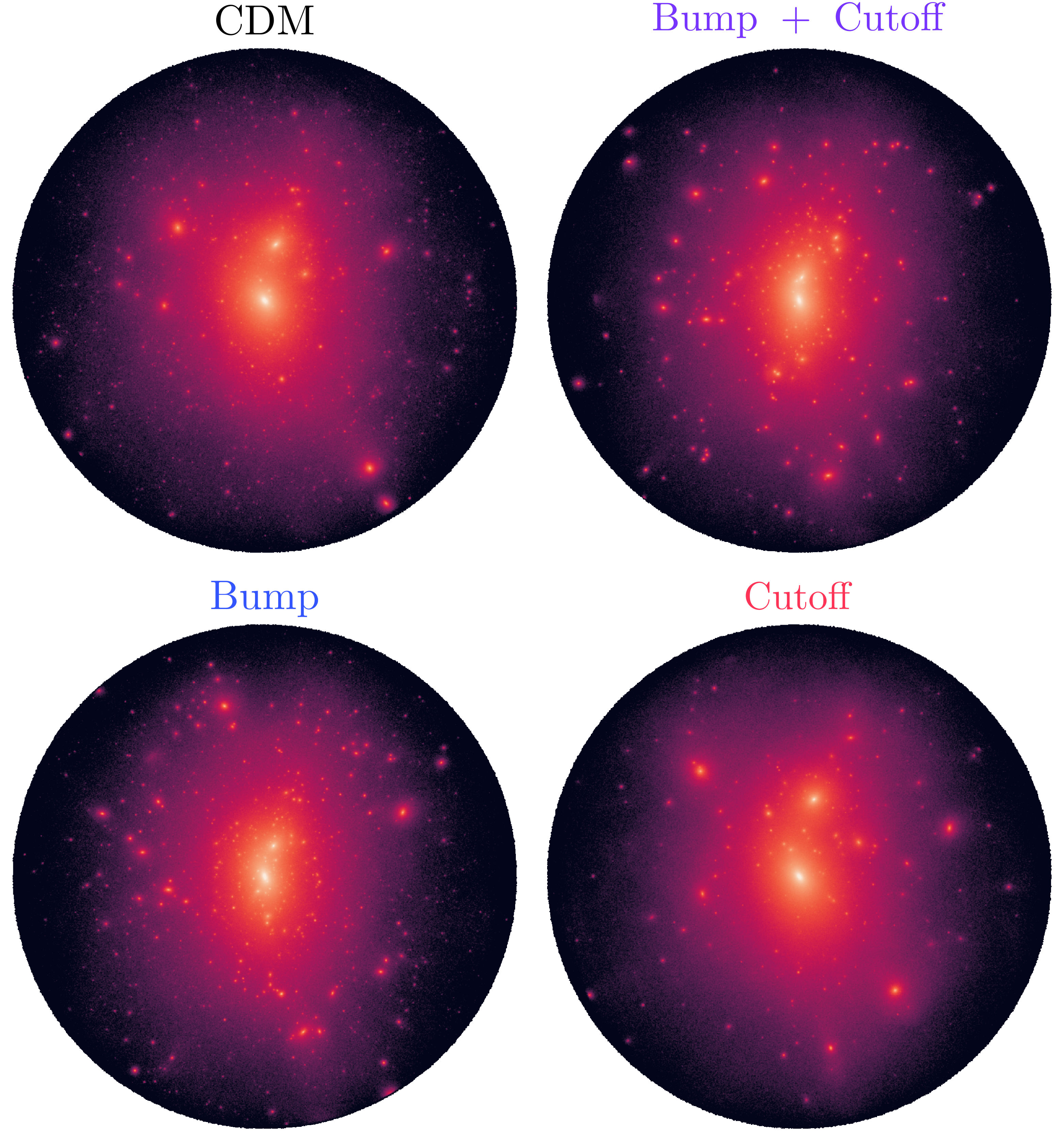}
    \caption{Projected DM density maps for our CDM (top left), Bump + Cutoff (top right), Bump (bottom left), and Cutoff (bottom right) simulations; for all non-CDM models, we show the $k_0=22.8~\mathrm{Mpc}^{-1}$ result. Each visualization is centered on the host, spans $1.5$ times its virial radius, and is created using \textsc{meshoid} (\url{https://github.com/mikegrudic/meshoid}).}
    \label{fig:vis_enhanced}
\end{figure*}

Host halo mass accretion histories in our zoom-ins with modified ICs are nearly identical to CDM. Figure~\ref{fig:vis_enhanced} shows density projections of the CDM, Bump + Cutoff, Bump, and Cutoff simulations with $k_0=22.8~\mathrm{Mpc}^{-1}$ at $z=0$. Subhalo abundances are visibly enhanced in the Bump + Cutoff and Bump simulations; in these scenarios, subhalos are found closer to the host center, and their inner densities are noticeably higher than in CDM. We quantify these results below.

\begin{figure}[t!]
\centering
\hspace{-5mm}
\includegraphics[trim={0 0.35cm 0 0cm},width=0.49\textwidth]{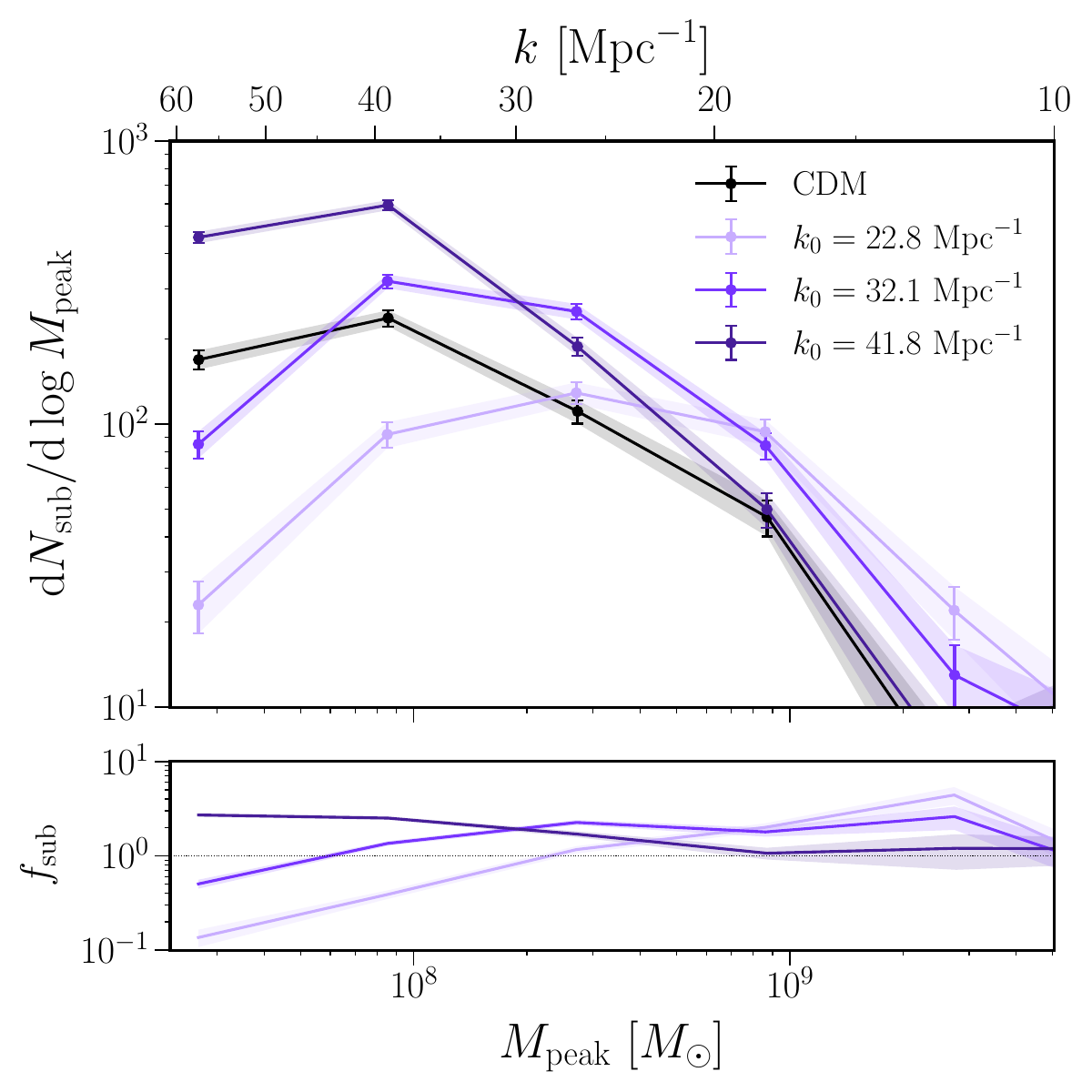}
    \caption{Differential SHMFs in CDM (black) and Bump + Cutoff (purple) models with $k_0=22.8$, $32.1$, and $41.8~\mathrm{Mpc}^{-1}$, from lightest to darkest shade. The bottom panel shows the ratio to CDM (Equation~\ref{eq:fsub}). In both panels, shaded bands show $1\sigma$ Poisson uncertainties.}
    \label{fig:enhanced_shmf}
\end{figure}


\section{Subhalo Mass Functions}
\label{sec:results}

Figure~\ref{fig:enhanced_shmf} shows differential SHMFs in CDM and our three Bump + Cutoff simulations at $z=0$. We restrict this figure to $M_{\mathrm{peak}}\lesssim 10^9~M_{\mathrm{\odot}}$ since all simulations contain $\lesssim 10$ subhalos, and are thus subject to large statistical uncertainties, above this threshold. We measure SHMFs down to $M_{\mathrm{peak}}=1.5\times 10^7~M_{\mathrm{\odot}}$, which coincides with our $M_{\mathrm{vir}}$ cut. Note that the CDM SHMF flattens near $M_{\mathrm{peak}}=1.5\times 10^7~M_{\mathrm{\odot}}$ because of this $M_{\mathrm{vir}}$ cut.

All Bump + Cutoff models enhance the SHMF over certain ranges of $M_{\mathrm{peak}}$, but the shape and amplitude of this enhancement differ as a function of $k_0$. For $k_0=22.8$ and $32.1~\mathrm{Mpc}^{-1}$, the enhancement is significant at high $M_{\mathrm{peak}}$ and weakens at lower masses, producing a feature in the SHMF that reflects the shape of $P(k)$. For example, in the $k_0=22.8~\mathrm{Mpc}^{-1}$ case, the SHMF inflects at $M_{\mathrm{peak}}\approx10^9~M_{\mathrm{\odot}}$; this mass corresponds to $k\approx 20~\mathrm{Mpc}^{-1}$ in linear theory~\citep{Nadler241003635}, placing it near the peak of the $k_0=22.8~\mathrm{Mpc}^{-1}$ transfer function. The SHMF inflection occurs at lower $M_{\mathrm{peak}}$ for  $k_0=32.1~\mathrm{Mpc}^{-1}$; this is expected because the feature in $P(k)$ occurs at a higher wavenumber in this case. Furthermore, SHMFs are suppressed relative to CDM at the lowest $M_{\mathrm{peak}}$ we consider for both $k_0=22.8$ and $32.1~\mathrm{Mpc}^{-1}$, since subhalos are resolved at scales where $P(k)$ is suppressed in these models.

On the other hand, the $k_0=41.8~\mathrm{Mpc}^{-1}$ SHMF is increasingly enhanced relative to CDM down to the lowest values of $M_{\mathrm{peak}}$ we consider. Note that $P(k)$ suppression begins at $k\approx 80~\mathrm{Mpc}^{-1}$ in this case, which maps to $M_{\mathrm{peak}}\approx 10^7~M_{\mathrm{\odot}}$. This mass is below our resolution limit; as a result, we do not detect the effects of $P(k)$ suppression on the SHMF for $k_0=41.8~\mathrm{Mpc}^{-1}$. This is consistent with the behavior of the SHMFs in the $k_0=22.8$ and $32.1~\mathrm{Mpc}^{-1}$ models, which respond in a fairly localized manner to features in $P(k)$. All of these results are robust as a function of simulation resolution, as demonstrated in Appendix~\ref{sec:convergence}.

\begin{figure}[t!]
\centering
\hspace{-5mm}
\includegraphics[trim={0 0.35cm 0 0cm},width=0.49\textwidth]{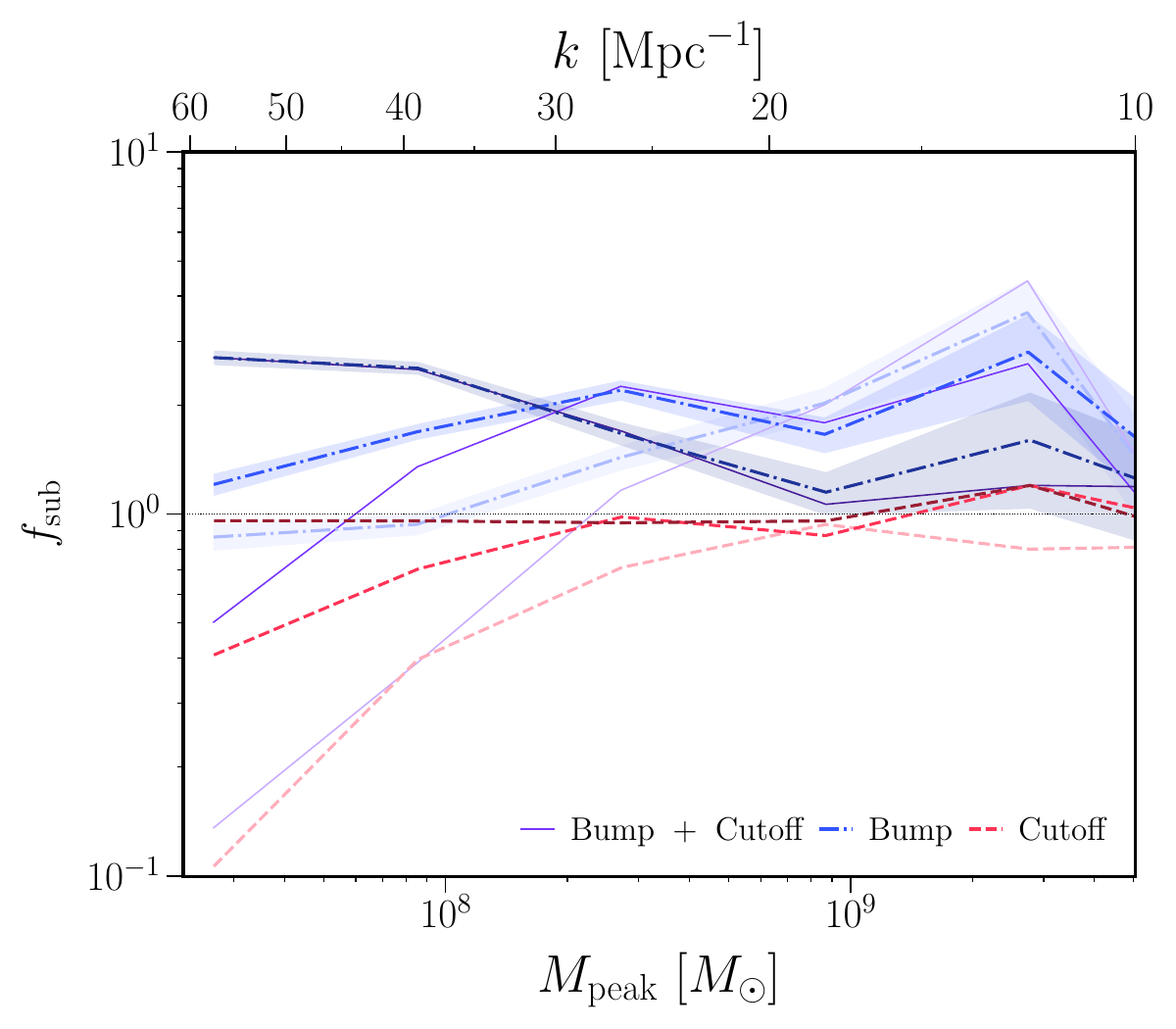}
    \caption{Ratio of the differential SHMF relative to CDM (Equation~\ref{eq:fsub}). Results are shown for Bump + Cutoff (solid purple), Bump (dotted--dashed blue), and Cutoff (dashed red) models with $k_0=22.8$, $32.1$, and $41.8~\mathrm{Mpc}^{-1}$, from lightest to darkest shade. Shaded bands show $1\sigma$ Poisson uncertainties for the Bump results.}
    \label{fig:sub_ratio}
\end{figure}

We define the differential SHMF ratio relative to CDM,
\begin{equation}
    f_{\mathrm{sub}} \equiv \frac{\left(\mathrm{d}N_{\mathrm{sub}}/\mathrm{d}\log M_{\mathrm{peak}}\right)_{P(k)}}{\left(\mathrm{d}N_{\mathrm{sub}}/\mathrm{d}\log M_{\mathrm{peak}}\right)_{\mathrm{CDM}}},\label{eq:fsub}
\end{equation}
where the numerator denotes either the Bump + Cutoff, Bump, or Cutoff scenario. Figure~\ref{fig:sub_ratio} shows $f_{\mathrm{sub}}$ for the Bump + Cutoff (solid), Bump (dotted--dashed), and Cutoff (dashed) scenarios. For both $k_0=22.8$ and $32.1~\mathrm{Mpc}^{-1}$, we find that the SHMF is more enhanced at small $M_{\mathrm{peak}}$ in the Bump versus the corresponding Bump + Cutoff model, while the corresponding SHMFs are nearly identical for $k_0=41.8~\mathrm{Mpc}^{-1}$. This behavior is due to differences in the underlying transfer functions, since we resolve the effects of $P(k)$ suppression for $k_0=22.8$ and $32.1~\mathrm{Mpc}^{-1}$, but not for $k_0=41.8~\mathrm{Mpc}^{-1}$, in the Bump + Cutoff scenario.

As shown in Figure~\ref{fig:sub_ratio}, SHMFs can differ even at scales where $P(k)$ is identical in the Bump + Cutoff and Bump scenarios. For example, $f_{\mathrm{sub}}$ for the $k_0=22.8~\mathrm{Mpc}^{-1}$ Bump model is larger than the corresponding Bump + Cutoff $f_{\mathrm{sub}}$ at $M_{\mathrm{peak}}=10^8~M_{\mathrm{\odot}}$, which maps to a wavenumber just above the $P(k)$ peak in these models (i.e., $k>k_0$). These differences indicate efficient small-to-large scale power transfer during nonlinear structure formation (e.g., via mergers). Small-to-large scale power transfer is generally thought to be a minor effect~\citep{Little1991,Bagla0408429,Bagla240805118} but it has not been studied in a zoom-in setting before.

As expected, SHMFs are strictly suppressed relative to CDM for all models in the Cutoff scenario. Moreover, Figure~\ref{fig:sub_ratio} provides means to compare the SHMFs arising in models where the ICs are perturbed away from CDM in opposite directions: the Cutoff case captures suppression, and the Bump + Cutoff case captures enhancement at the corresponding scale. We find that the amplitude of the suppression in Cutoff SHMFs is comparable to the amplitude of the enhancement in the Bump + Cutoff SHMFs, indicating that the amplitude of the effect does not depend on the sign of the perturbation to the ICs. These results are consistent with negligible differences in large-to-small-scale power transfer among our simulations. This power transfer direction is well studied (e.g., \citealt{Peebles1980,Bagla9605202}), but likely does not vary among our simulations because we fix the large-scale zoom-in environment across all runs. 

In Appendix~\ref{sec:cutoff_comparison}, we show that SHMF suppression is less severe in Cutoff models compared to the corresponding WDM models. This follows because WDM $P(k)$ suppression begins at larger scales compared to the Cutoff scenario, and confirms the finding from \cite{Nadler241003635} that SHMF suppression in WDM zoom-ins is largely determined by wavenumbers near the onset of the WDM transfer function suppression.

Although we simulate only one MW-like system, we expect all results throughout this section to be robust across different environments. In particular, \cite{Nadler241003635} demonstrated that $P(k)$ suppression leads to SHMF suppression relative to CDM that does not significantly vary from host to host. Nonetheless, including additional simulations would decrease the statistical uncertainty of our results.

\section{Physical Effects}
\label{sec:drivers}

\subsection{Isolated versus Subhalo Mass Function Ratios}
\label{sec:isolated_hmf}

\begin{figure}[t!]
\centering
\hspace{-5mm}
\includegraphics[trim={0 0.35cm 0 0cm},width=0.49\textwidth]{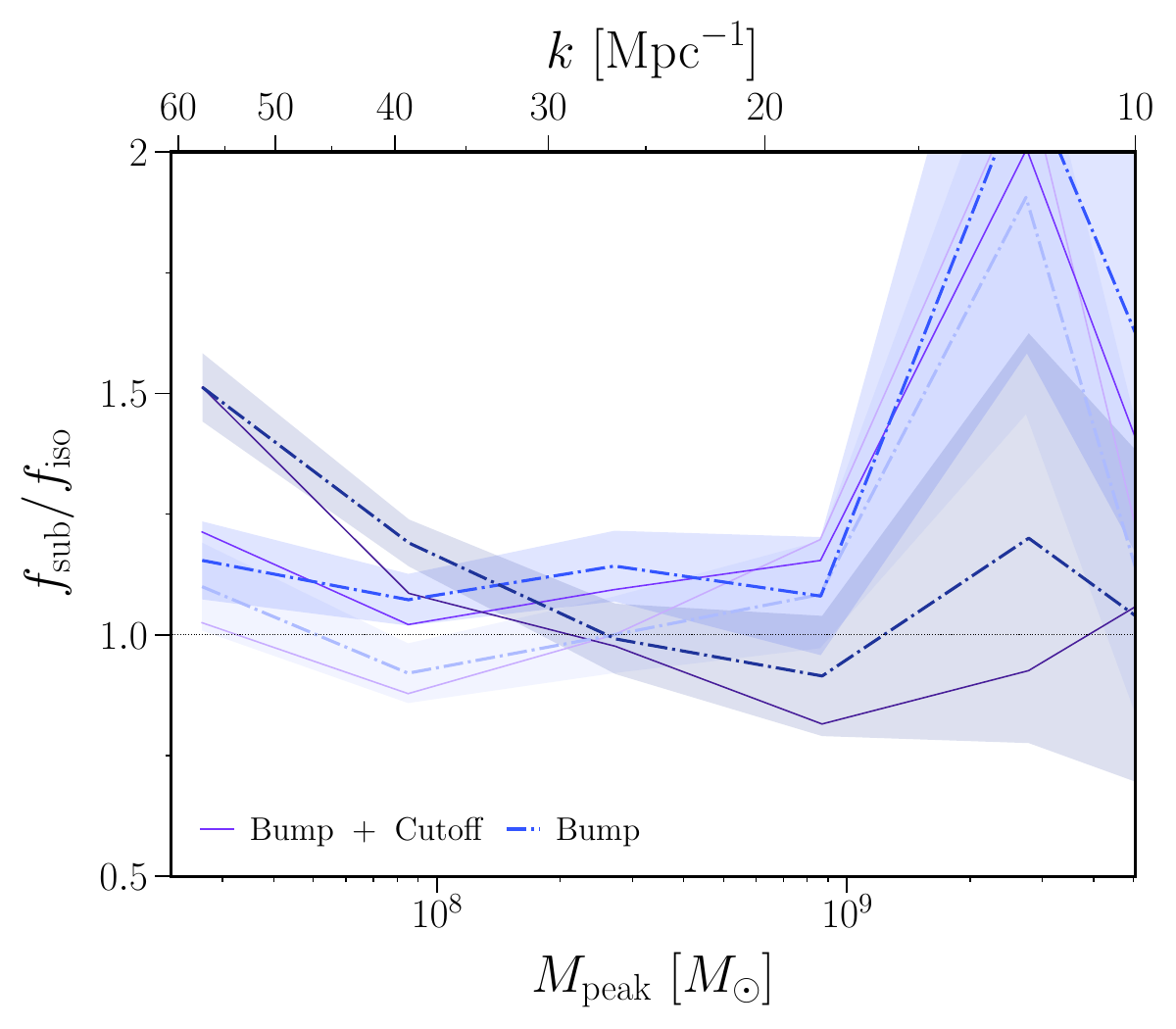}
    \caption{Ratio of the differential SHMF in our models with $P(k)$ enhancement relative to CDM divided by the corresponding ratio for isolated halos (Equation~\ref{eq:fiso}). Results are shown for Bump + Cutoff (solid purple) and Bump (dotted--dashed blue) models with $k_0=22.8$, $32.1$, and $41.8~\mathrm{Mpc}^{-1}$, from lightest to darkest shade. Shaded bands show $1\sigma$ Poisson uncertainties for the Bump results. The differences in subhalo vs.\ isolated halo abundances shown here are small compared to the differences in SHMFs relative to CDM that are imprinted by $P(k)$ (see Figure~\ref{fig:sub_ratio}).}
    \label{fig:sub_iso_ratio}
\end{figure}

SHMFs in our models with $P(k)$ enhancement may deviate from CDM due to changes in the infall mass function or differences in tidal evolution. To test these explanations, we define isolated halos as objects that have never been a subhalo of a larger host and are within $2.63~\mathrm{Mpc}$ of the host center (i.e., 10 times the virial radius of the MW host) at $z=0$.\footnote{This definition of isolated halos excludes splashback systems that have previously orbited within a larger host.} We then measure the ratio of the mass function for halos in the zoom-in region that are isolated at $z=0$ relative to CDM,
\begin{equation}
    f_{\mathrm{iso}} \equiv \frac{\left(\mathrm{d}N_{\mathrm{iso}}/\mathrm{d}\log M_{\mathrm{peak}}\right)_{P(k)}}{\left(\mathrm{d}N_{\mathrm{iso}}/\mathrm{d}\log M_{\mathrm{peak}}\right)_{\mathrm{CDM}}}\label{eq:fiso}
\end{equation}
Figure~\ref{fig:sub_iso_ratio} shows $f_{\mathrm{sub}}/f_{\mathrm{iso}}$ in our Bump + Cutoff and Bump models. We omit the Cutoff models since the $k_0=22.8~\mathrm{Mpc}^{-1}$ case behaves similarly to the corresponding Bump + Cutoff model at low $M_{\mathrm{peak}}$, while the results for higher-$k_0$ models are consistent with unity at all $M_{\mathrm{peak}}$ that we consider.

To first order, Figure~\ref{fig:sub_iso_ratio} shows that changes to the SHMF are determined by changes to the isolated HMF. In particular, $f_{\mathrm{sub}}/f_{\mathrm{iso}}$ differs from unity by at most $\approx 25\%$ in all models, except at the high-$M_{\mathrm{peak}}$ end, which we discuss below. Although this difference is significant given the statistical errors on our SHMF measurements, it is small relative to the variations in $f_{\mathrm{sub}}$ shown in the bottom panel of Figure~\ref{fig:enhanced_shmf}.

The behavior of $f_{\mathrm{sub}}/f_{\mathrm{iso}}$ for the $k_0=22.8~\mathrm{Mpc}^{-1}$ Bump + Cutoff model is particularly informative. In this model, both $P(k)$ enhancement and suppression are well resolved. The scale of the $P(k)$ enhancement corresponds to the highest $M_{\mathrm{peak}}$ shown in Figure~\ref{fig:sub_iso_ratio}, while the $P(k)$ suppression scale corresponds to the lowest $M_{\mathrm{peak}}$ we plot. At high $M_{\mathrm{peak}}$, $f_{\mathrm{sub}}/f_{\mathrm{iso}}$ becomes much larger than unity, indicating that subhalos formed from modes near the $P(k)$ peak are more resilient to tidal disruption than CDM subhalos with similar values of $M_{\mathrm{peak}}$, and vice versa for subhalos with low $M_{\mathrm{peak}}$.\footnote{Note that $M_{\mathrm{peak}}$ itself does not significantly differ among matched (sub)halos in the different simulations.} This is consistent with the effects of $P(k)$ on subhalo concentrations demonstrated in Section~\ref{sec:m-c}. Furthermore, the differences between the Bump + Cutoff and Bump results are consistent with this explanation. For example, at low $M_{\mathrm{peak}}$, $f_{\mathrm{sub}}/f_{\mathrm{iso}}$ is not significantly suppressed in the Bump model since it does not have a $P(k)$ cutoff. Meanwhile, the $P(k)$ peak in the $k_0=41.8~\mathrm{Mpc}^{-1}$ case corresponds to the lowest $M_{\mathrm{peak}}$ we plot, which explains why $f_{\mathrm{sub}}/f_{\mathrm{iso}}$ rises toward low $M_{\mathrm{peak}}$ in this case, as does $f_{\mathrm{sub}}$ itself.

\subsection{Subhalo Radial Distribution}
\label{sec:orbit}

The changes to the SHMF in our models with $P(k)$ enhancement have implications for the subhalo radial distribution. In particular, subhalos with higher infall masses sink to the host center more quickly due to dynamical friction, resulting in a more concentrated radial distribution (e.g., \citealt{Nadler220902675}). Thus, we expect a more centrally concentrated radial distribution for models that enhance the SHMF relative to CDM at high $M_{\mathrm{peak}}$, and a less centrally concentrated radial distribution for models that enhance the SHMF at low $M_{\mathrm{peak}}$. Conversely, we expect models that suppress the SHMF at low $M_{\mathrm{peak}}$ to yield a more centrally concentrated radial distribution (e.g., \citealt{Lovell210403322}).

\begin{figure}[t!]
\centering
\hspace{-4mm}
\includegraphics[trim={0 0.35cm 0 0cm},width=0.49\textwidth]{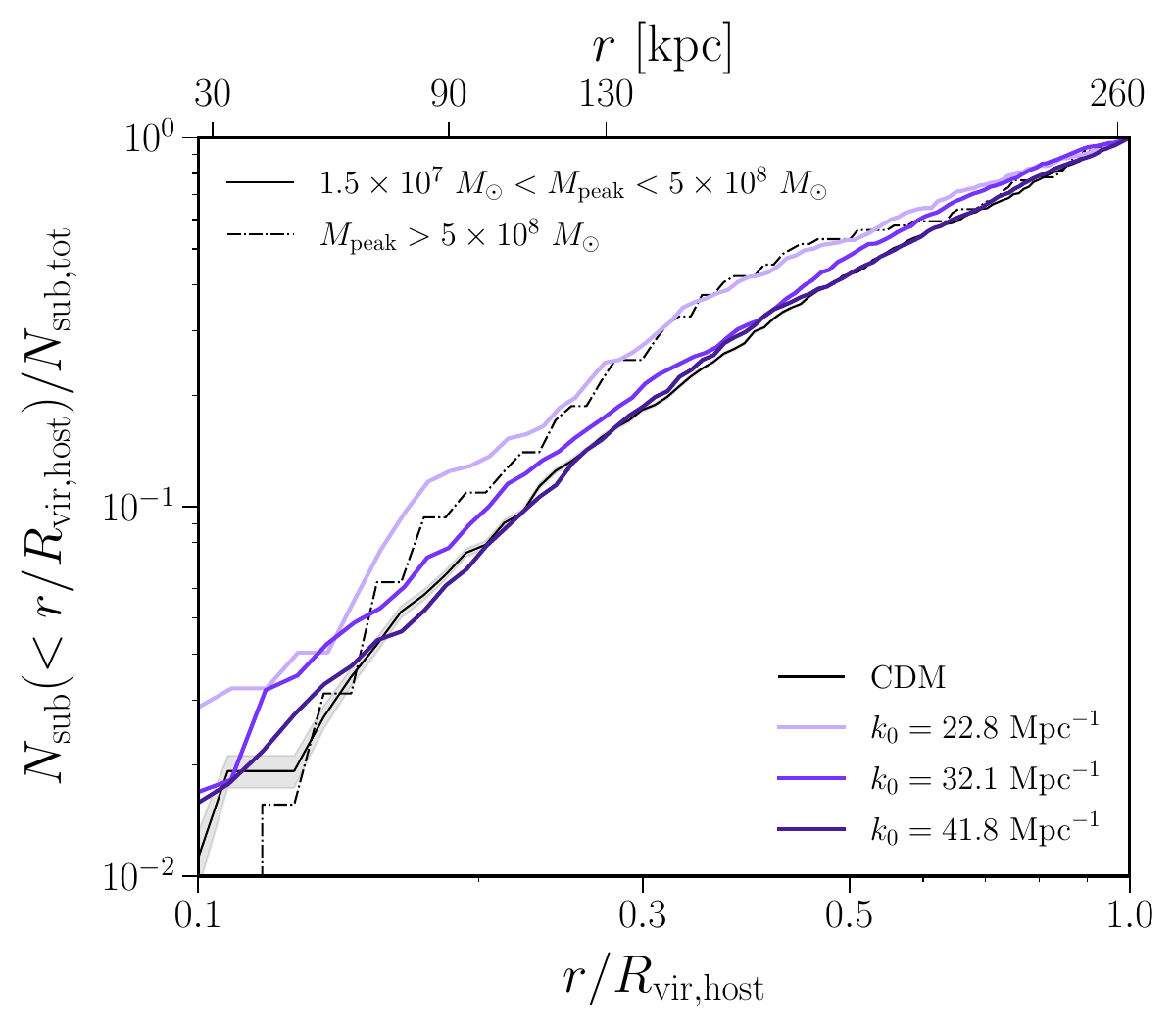}
    \caption{Normalized radial distribution of subhalos with $1.5\times 10^7~M_{\mathrm{\odot}}<M_{\mathrm{peak}}<5\times 10^8~M_{\mathrm{\odot}}$ in CDM (black) and Bump + Cutoff models (purple) with $k_0=22.8$, $32.1$, and $41.8~\mathrm{Mpc}^{-1}$, from lightest to darkest shade. The dotted--dashed black line shows the CDM result for subhalos with $M_{\mathrm{peak}}>5\times 10^8~M_{\mathrm{\odot}}$. The shaded gray band shows $1\sigma$ Poisson uncertainty for the CDM result. Note that the host halo's virial radius is $R_{\mathrm{vir,host}}=263~\mathrm{kpc}$ in all simulations. Models that enhance $P(k)$ at higher $M_{\mathrm{peak}}$ yield more centrally concentrated radial distributions.}
    \label{fig:enhanced_radial}
\end{figure}

Figure~\ref{fig:enhanced_radial} shows normalized radial distributions of subhalos with $1.5\times 10^7~M_{\mathrm{\odot}}<M_{\mathrm{peak}}<5\times 10^8~M_{\mathrm{\odot}}$ in CDM and Bump + Cutoff scenario, where the lower limit corresponds to our resolution cut. Note that we only show the Bump + Cutoff results for visual clarity; we discuss the other scenarios below. We find that the Bump + Cutoff radial distributions become more centrally concentrated as $k_0$ decreases; for example, the $k_0=22.8~\mathrm{Mpc}^{-1}$ result for $1.5\times 10^7~M_{\mathrm{\odot}}<M_{\mathrm{peak}}<5\times 10^8~M_{\mathrm{\odot}}$ is similar to the CDM result for $M_{\mathrm{peak}}>5\times 10^8~M_{\mathrm{\odot}}$. We also illustrate the mass dependence of the radial distribution by overplotting the CDM radial distribution for subhalos with $M_{\mathrm{peak}}>5\times 10^8~M_{\mathrm{\odot}}$.\footnote{This threshold is chosen to ensure that there are enough high-mass subhalos to reliably measure the radial distribution when splitting the population in this way.} Higher-$M_{\mathrm{peak}}$ subhalos are significantly more centrally concentrated in CDM, illustrating that the effects of $P(k)$ enhancement are partially degenerate with the standard mass dependence of the radial distribution in CDM.

We posit that two main effects drive these results. In the Bump + Cutoff scenario, the abundance of high-mass subhalos is enhanced relative to CDM, while the abundance of low-mass subhalos is suppressed. In our simulations, the radial distribution of higher-mass subhalos is expected to be more centrally concentrated than relative to lower-mass subhalos~\citep{Nadler220902675}. Thus, these effects combine to produce an overall enhancement of the inner radial distribution. We have checked that the individual effects of $P(k)$ enhancement (in the Bump scenario) and suppression (in the Cutoff scenario) affect the radial distribution less strongly than in the Bump + Cutoff scenario, which supports this reasoning. Another potential explanation is that subhalo mass-loss rates are mass dependent on the mass scales corresponding to wavenumbers where $P(k)$ is altered relative to CDM (e.g., \citealt{He230901109}). However, \cite{Nadler241003635} found that WDM mass-loss rates do not significantly differ from CDM in simulations with similar cutoff scales and resolution compared to those we analyze here. We leave a dedicated study of subhalo mass-loss rates in enhanced $P(k)$ models to future work.

Finally, we note that matched massive subhalos are found closer to the host center in models with $P(k)$ enhancement, and farther from the host center in models with $P(k)$ suppression, relative to CDM. Figure~\ref{fig:vis_enhanced} illustrates this for the LMC analog---the largest subhalo directly above the host center---in the Bump + Cutoff, Bump, and Cutoff scenarios. These LMC analogs' orbits and infall times are nearly identical before infall, but in the enhanced scenarios, they sink to the host center more quickly after crossing into the host's virial radius. This could be due to differences in mass-loss rates, as discussed above, or due to more efficient dynamical friction in enhanced $P(k)$ scenarios; the latter effect is also worth a dedicated study.

\subsection{Subhalo Mass--Concentration Relation}
\label{sec:m-c}

We expect (sub)halo concentrations to increase relative to CDM in scenarios with $P(k)$ enhancement, and decrease in scenarios with $P(k)$ suppression. We quantify this effect using subhalos with $\geq 2000$ particles, or $M_{\mathrm{vir}}>10^8~M_{\mathrm{\odot}}$, using the effective concentration~\citep{Yang221113768}
\begin{equation}
    c_{\mathrm{eff}} = \frac{R_{\mathrm{vir}}}{R_{\mathrm{max}}/2.126},\label{eq:c_eff}
\end{equation}
where $R_{\mathrm{vir}}$ is the virial radius and $R_{\mathrm{max}}$ is the radius from the center of a given subhalo at which the maximum circular velocity $V_{\mathrm{max}}$ is achieved. We use $c_{\mathrm{eff}}$ because it reduces to the virial concentration for Navarro--Frenk--White (NFW; \citealt{Navarro1997}) halos while flexibly capturing non-NFW profiles resulting from changes to $P(k)$ and/or tidal stripping, as well as from dynamical DM signatures like self-interactions~\citep{Yang221113768}.

\begin{figure}[t!]
\centering
\hspace{-4mm}
\includegraphics[trim={0 0.35cm 0 0cm},width=0.49\textwidth]{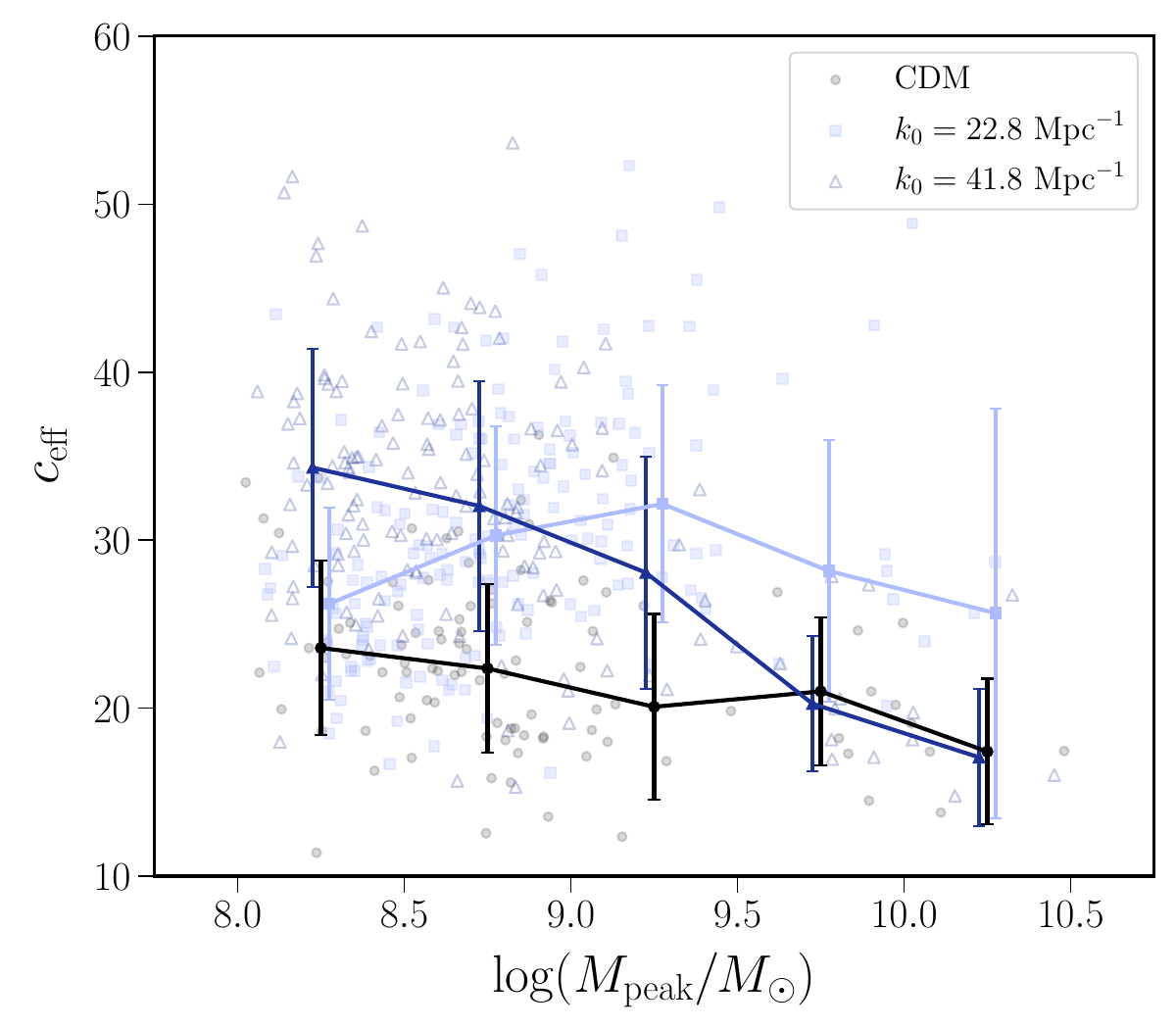}
    \caption{Effective concentration (Equation~\ref{eq:c_eff}) versus peak virial mass for subhalos with $M_{\mathrm{vir}}>10^8~M_{\mathrm{\odot}}$ in our CDM simulation (black circles), along with our $k_0=22.8~\mathrm{Mpc}^{-1}$ (light blue squares) and $41.8~\mathrm{Mpc}^{-1}$ (dark blue triangles) Bump simulations. Thick points show median values, and error bars show $1\sigma$ scatter in bins of $M_{\mathrm{peak}}$. Subhalo concentrations are enhanced in the Bump and Bump + Cutoff scenarios.}
    \label{fig:m_c}
\end{figure}

Figure~\ref{fig:m_c} shows the $c_{\mathrm{eff}}$--$M_{\mathrm{peak}}$ relation for subhalos in our CDM and $k_0=22.8$ and $41.8~\mathrm{Mpc}^{-1}$ Bump simulations. Subhalos in the Bump simulations are generally more concentrated than their CDM counterparts, reaching $c_{\mathrm{eff}}\approx 50$, whereas the most concentrated CDM subhalos have $c_{\mathrm{eff}}\approx 30$. This enhancement is most pronounced at $M_{\mathrm{peak}}\approx 10^{9.5}~M_{\mathrm{\odot}}$ in the $k_0=22.8~\mathrm{Mpc}^{-1}$ case and at $M_{\mathrm{peak}}\approx 10^8~M_{\mathrm{\odot}}$ in the $k_0=41.8~\mathrm{Mpc}^{-1}$. This reflects the shift of the peak in the underlying transfer function between the two scenarios and aligns with the mass scales where the respective SHMFs are most strongly enhanced relative to CDM. The concentration distribution for the $k_0=32.1~\mathrm{Mpc}^{-1}$ model peaks at intermediate masses, as expected.

We find that this behavior is similar in the Bump + Cutoff scenario, although the high-$c_{\mathrm{eff}}$ region of Figure~\ref{fig:m_c} is somewhat less populated in these models. Meanwhile, subhalos in the Cutoff runs systematically shift toward smaller $c_{\mathrm{eff}}$ compared to CDM due to $P(k)$ suppression, consistent with the results in \cite{Nadler241213065} for simulations of self-interacting DM models with $P(k)$ suppression but without late-time scattering. Across all $P(k)$ scenarios and values of $k_0$, we find that differences in the $c_{\mathrm{eff}}$--$M_{\mathrm{peak}}$ relation relative to CDM are similar when considering isolated halos rather than subhalos.

The response of the mass--concentration relation to $P(k)$ enhancement and/or suppression can thus be used to place constraints that are complementary to---but not entirely independent of---constraints from (sub)halo abundances. This idea has recently been used to constrain cosmologies where $P(k)$ is affected over a range of scales using dwarf galaxy stellar velocity dispersions~\citep{Esteban230604674,Dekker240704198} and strong-lensing substructure~\citep{Gilman211203293}. It will be interesting to revisit this signature to probe models with localized features in $P(k)$.

\section{Summary and Discussion}
\label{sec:discussion}

\subsection{Summary}

We have presented new zoom-in simulations of an MW analog with linear matter power spectrum enhancement and/or suppression. We highlight the following key results:
\begin{itemize}
    \item The subhalo mass function reflects the shape of $P(k)$, including both its enhancement and suppression when they are present (Figures~\ref{fig:enhanced_shmf} and \ref{fig:sub_ratio});
    \item Changes to the SHMF relative to CDM are mainly imprinted at infall, rather than during tidal evolution (Figure~\ref{fig:sub_iso_ratio});
    \item $P(k)$ enhancement yields more centrally concentrated subhalo radial distributions (Figure~\ref{fig:enhanced_radial}).
\end{itemize}

These results are consistent with previous simulations that include Gaussian $P(k)$ enhancement, but probe a different range of spatial and temporal scales. In particular, \cite{Tkachev230713774,Tkachev240702991}, \cite{Pilipenko240417803}, and \cite{Eroshenko240902739} found that a Gaussian $P(k)$ bump leads to HMF enhancement relative to CDM at high redshifts. Figures~\ref{fig:enhanced_shmf} and \ref{fig:sub_ratio} demonstrate that this enhancement persists in the $z=0$ SHMF. This finding is not guaranteed from the high-$z$ results, since mode coupling tends to erase power spectrum features over time. These previous studies also showed that central densities of halos in the Gaussian enhancement scenarios are increased relative to CDM, consistent with Figure~\ref{fig:m_c}.

Our results are also qualitatively consistent with high-resolution simulations of blue-tilted $P(k)$ models~\citep{Wu241216072}, including with PBHs~\citep{Colazo250523896}. In particular, both of these studies found that (sub)halo abundances are enhanced at a mass scale that maps to the wavenumber of the $P(k)$ enhancement, and \cite{Wu241216072} also found that subhalos in blue-tilted models are more centrally concentrated and have enhanced $V_{\mathrm{max}}$ values relative to CDM.

\subsection{Implications for Semianalytic Models}

The relatively tight correspondence that we find between $P(k)$ and SHMF features has important consequences for semianalytic structure formation models. In scenarios with $P(k)$ suppression, extended Press--Schechter (ePS) window functions that truncate sharply as a function of wavenumber predict mass functions more accurately compared to sharp real-space filters, which work well in CDM (e.g., \citealt{Benson12093018,Leo180102547}). This situation is relatively unexplored for models with $P(k)$ enhancement, but our results generally suggest that the mapping from $P(k)$ to mass functions remains relatively sharp in these scenarios, including when $P(k)$ suppression is also present.

We note that ePS models can make surprising predictions in scenarios with $P(k)$ enhancement. For example, \cite{Balaji240811098} predicted that strict $P(k)$ enhancement relative to CDM in a supermassive PBH model can lead to SHMF suppression at masses that correspond to scales smaller than the $P(k)$ enhancement ($k>k_0$). Mathematically, this follows from combining a Gaussian $P(k)$ enhancement with an exponential $k$-space ePS window function. Physically, these authors reasoned that high-mass halos formed from modes near the transfer function peak collapse earlier than in CDM, preventing the formation of smaller halos within these systems. We do not find evidence of this effect in our Bump simulations down to the smallest values of $M_{\mathrm{peak}}$ we resolve, as shown by Figure~\ref{fig:sub_ratio}. Higher-resolution zoom-in simulations would be needed to study whether this effect is relevant at lower values of $M_{\mathrm{peak}}$.

\subsection{Future Work}

$P(k)$ enhancement and suppression can arise in many nonstandard inflation and DM models. For example, features in the inflaton potential imprint oscillations in the primordial power spectrum, which can translate to $P(k)$ enhancements that qualitatively resemble the models we simulate~\citep{Adams0102236}. Meanwhile, axions produced with a large misalignment angle yield $P(k)$ that resemble our Bump + Cutoff transfer function model~\citep{Arvanitaki190911665}. We expect that our simulation results can be mapped to these scenarios. The opposite behavior---i.e., a $P(k)$ cutoff followed on smaller scales by an enhancement---can arise in ultralight dark matter models due to a combination of free-streaming and white-noise effects~\citep{Amin221109775}. The mapping from our Gaussian bump scenario to specific models will be addressed in future work.

Additional simulations will be needed to capture DM substructure in models that affect $P(k)$ differently than our fiducial Gaussian bump model. These include cosmologies with a running spectral index (e.g., following \citealt{Garrison-Kimmel14053985} but using a positive rather than negative running), or a blue-tilted enhancement (e.g., \citealt{Wu241216072}); the latter scenario can also include a PBH population (e.g., \citealt{Colazo250523896}). This is a timely direction for future work because these models have recently been constrained using small-scale structure observables~\citep{Gilman211203293,Balaji240811098,Esteban230604674,Dekker240704198}.

\subsection{Outlook} 

Our results underscore the importance of modeling the mapping from $P(k)$ to small-scale DM structure using cosmological simulations. In particular, (sub)halo populations sourced from density perturbations over a given range of wavenumbers are sensitive to the shape of $P(k)$ due to nonlinear mode coupling during structure formation. Thus, our results affirm optimism that $P(k)$ reconstruction is feasible on small scales, following classic techniques developed for large-scale structure observables (e.g., \citealt{Weinberg1992}). 

The simulations presented in this work will help enable $P(k)$ reconstruction using the Milky Way satellite population. This is an exciting direction given upcoming data from the Vera C.\ Rubin Observatory, which is expected to significantly increase the number of detected satellites~\citep{Tsiane240416203} and improve DM constraints~\citep{BechtolLSST,Drlica-Wagner190201055}. In particular, \cite{Nadler240110318} showed that SHMF enhancement can be probed using upcoming satellite luminosity function observations; in turn, these constraints can be mapped to $P(k)$. Complementary data, including stellar velocity dispersion measurements, will likely further boost this sensitivity, although hydrodynamic simulations in enhanced $P(k)$ scenarios will be needed to assess this precisely. More generally, deriving robust $P(k)$ constraints using DM substructure will require the development of a fast, flexible, and general forward modeling framework, which will be the focus of future studies.


\section*{Acknowledgments}

Halo catalogs, merger trees, and particle snapshots are distributed in Zenodo at \url{https://zenodo.org/records/16915369}~\citep{Nadler2025COZMIC}. Analysis code is available at \url{https://github.com/eonadler/COZMIC/}.

This material is based upon work supported by the National Science Foundation under grant No.\ 2509561 (E.O.N.\ and A.B.) and No.\ 2407380 (V.G.). Any opinions, findings, and conclusions or recommendations expressed in this material are those of the author(s) and do not necessarily reflect the views of the NSF. This research was supported in part by grant NSF PHY-2309135 to the Kavli Institute for Theoretical Physics (KITP). V.G.\ additionally acknowledges the support from NASA through
the Astrophysics Theory Program, Award No.\ 21-ATP21-0135, from the National Science Foundation (NSF) CAREER grant No.\ PHY-2239205, and from the Research Corporation for Science Advancement through the Cottrell Scholars program.

The computations presented here were conducted through
Carnegie’s partnership in the Resnick High Performance
Computing Center, a facility supported by Resnick Sustainability Institute at Caltech. This work used data from COZMIC I, which is available at \href{https://zenodo.org/records/14649137}{10.5281/zenodo.14649137}. This work used data from the Milky Way-est suite of simulations, hosted at \href{https://web.stanford.edu/group/gfc/gfcsims/}{https://web.stanford.edu/group/gfc/gfcsims/}, which was supported by the
Kavli Institute for Particle Astrophysics and Cosmology at
Stanford, SLAC National Accelerator Laboratory, and the
U.S. DoE under contract number DE-AC02-76SF00515 to
SLAC.

\software{
{\sc consistent-trees} \citep{Behroozi11104370},
\textsc{Helpers} (\http{bitbucket.org/yymao/helpers/src/master/}),
\textsc{Jupyter} (\http{jupyter.org}),
\textsc{Matplotlib} \citep{matplotlib},
\textsc{NumPy} \citep{numpy},
\textsc{pynbody} \citep{pynbody},
{\sc Rockstar} \citep{Behroozi11104372},
\textsc{SciPy} \citep{scipy},
\textsc{Seaborn} (\https{seaborn.pydata.org}).
}

\bibliographystyle{yahapj2}
\bibliography{references,software}


\appendix

\section{Convergence Tests}
\label{sec:convergence}

To test for convergence, we resimulate all models using one fewer \textsc{MUSIC} refinement region when generating ICs, which results in a high-resolution DM mass of $4\times 10^5~M_{\mathrm{\odot}}$. For these low-resolution (LR) runs, we use a softening of $\epsilon=170~\mathrm{pc}~h^{-1}$. Large-scale features of the simulations, including the host halo mass accretion history, are nearly identical between resolution levels. In the following analyses, we apply a $M_{\mathrm{vir}}>1.2\times 10^8~M_{\mathrm{\odot}}$ cut to the resulting subhalo populations, which corresponds to $300$ particles in the LR runs.

\begin{figure}[t!]
\centering
\hspace{-5mm}
\includegraphics[trim={0 0.35cm 0 0cm},width=0.49\textwidth]{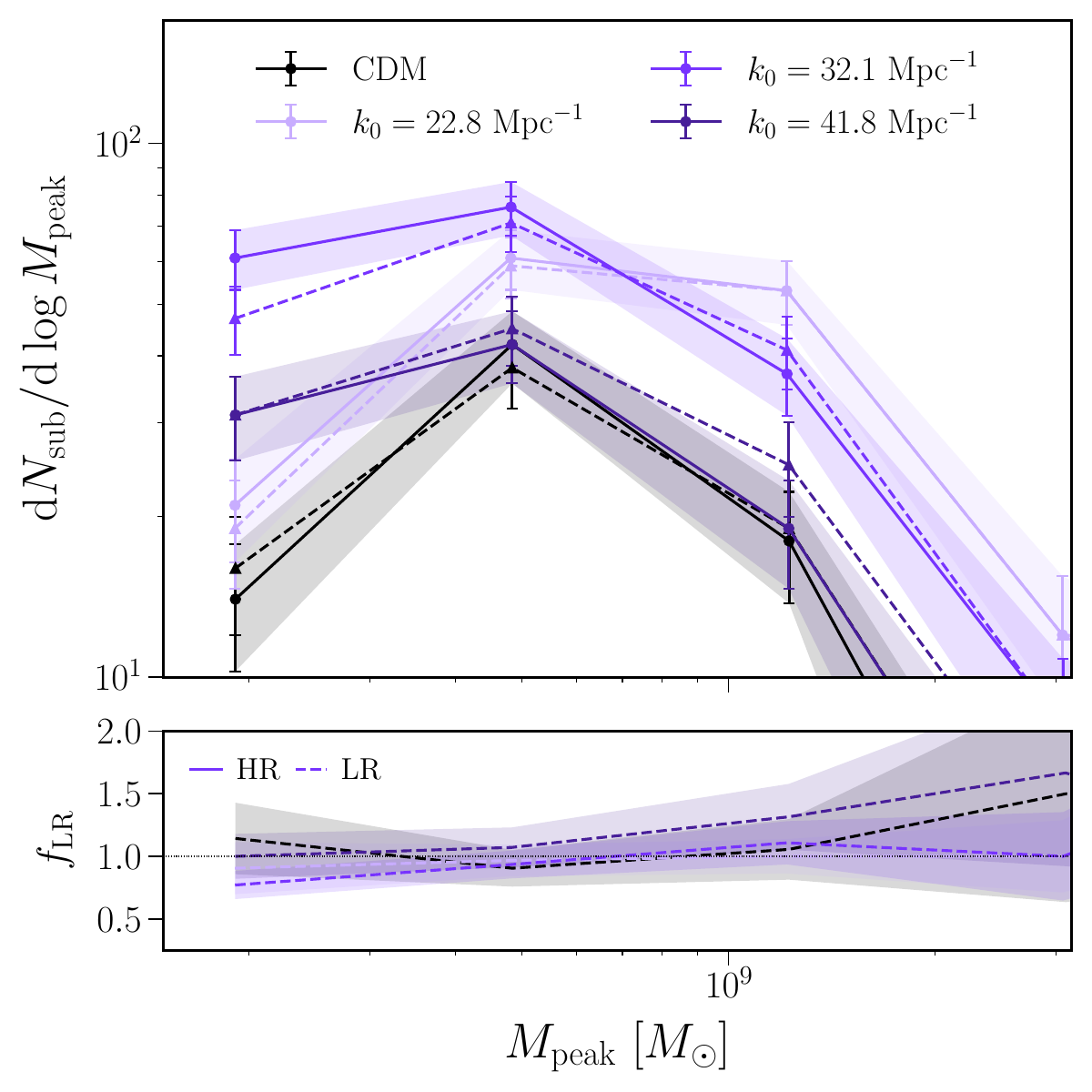}
    \caption{Same as Figure~\ref{fig:enhanced_shmf}, but comparing the differential SHMF in our LR (dashed) vs.\ HR (solid) simulations. We show CDM (black) and Bump + Cutoff (purple) models with $k_0=22.8$, $32.1$, and $41.8~\mathrm{Mpc}^{-1}$, from lightest to darkest shade. The bottom panel shows the SHMF ratio between each LR/HR pair. In both panels, shaded bands show $1\sigma$ Poisson uncertainties for the HR results.}
    \label{fig:shmf_convergence}
\end{figure}

The top panel of Figure~\ref{fig:shmf_convergence} compares the differential SHMF in the CDM and Bump + Cutoff LR runs to the high-resolution (HR) simulations we used in our main analysis. In the bottom panel, we further compare the runs using
\begin{equation}
    f_{\mathrm{LR}} \equiv \frac{\left(\mathrm{d}N_{\mathrm{sub,LR}}/\mathrm{d}\log M_{\mathrm{peak}}\right)_{P(k)}}{\left(\mathrm{d}N_{\mathrm{sub,HR}}/\mathrm{d}\log M_{\mathrm{peak}}\right)_{P(k)}}.
\end{equation}
All $f_{\mathrm{LR}}$ measurements are consistent with unity, given the Poisson uncertainties on our LR SHMFs. Thus, our SHMF measurements are converged down to a $300$-particle limit, lending confidence to our main results. We obtain similar levels of agreement for the Bump and Cutoff scenarios.

\section{SHMF Comparison for Cutoff versus WDM Scenarios}
\label{sec:cutoff_comparison}

\begin{figure}[t!]
\centering
\hspace{-5mm}
\includegraphics[trim={0 0.35cm 0 0cm},width=0.49\textwidth]{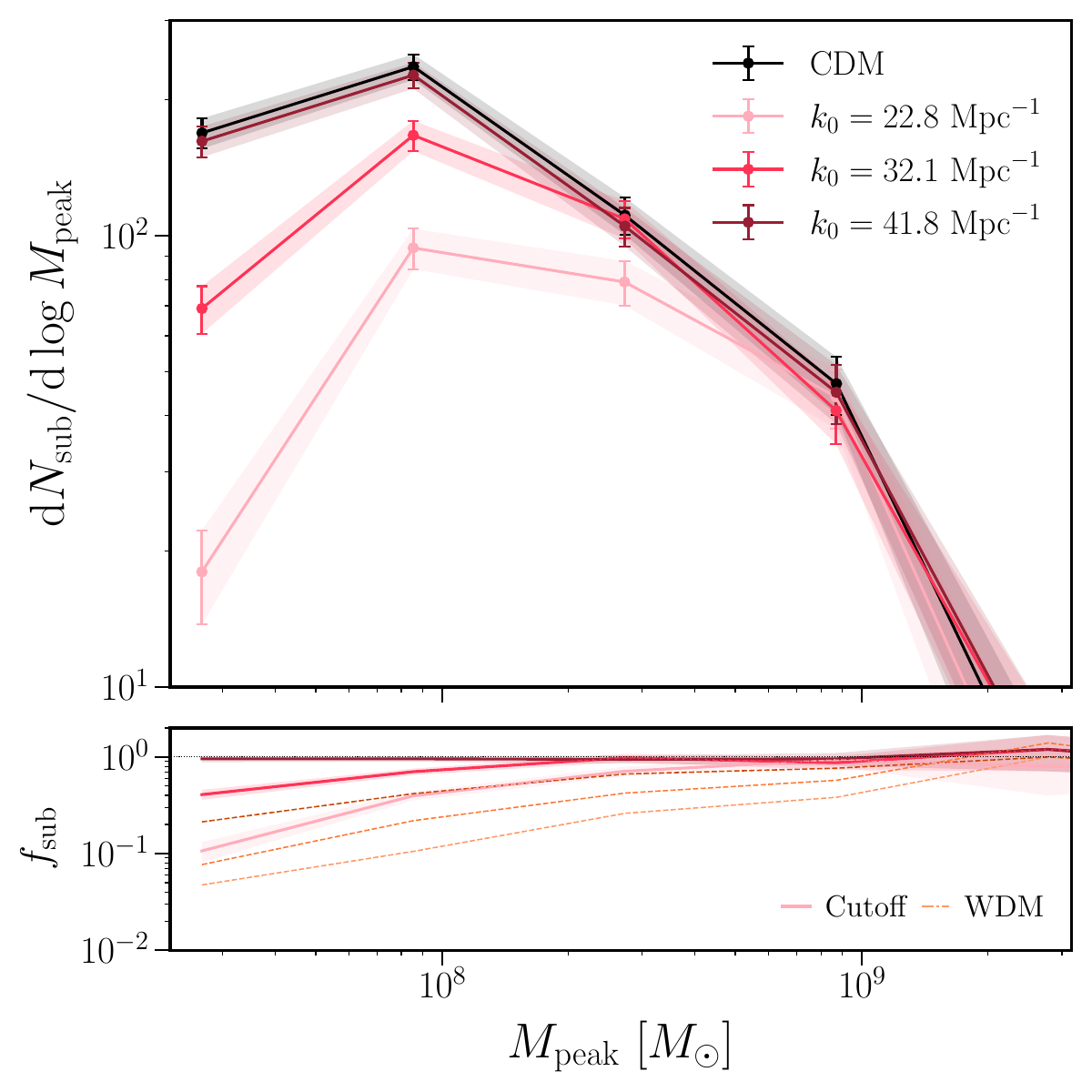}
    \caption{Same as Figure~\ref{fig:enhanced_shmf}, but comparing the differential SHMF in CDM (black) to Cutoff (red) models with $k_0=22.8$, $32.1$, and $41.8~\mathrm{Mpc}^{-1}$, from lightest to darkest shade. The bottom panel shows the ratio of the SHMF to CDM in the Cutoff (solid red) and WDM (dashed orange) scenarios. In both panels, shaded bands show $1\sigma$ Poisson uncertainties.}
    \label{fig:suppressed_shmf}
\end{figure}

Figure~\ref{fig:suppressed_shmf} compares the SHMF in the Cutoff and WDM scenarios. As discussed in Section~\ref{sec:results}, WDM SHMFs are more suppressed than the corresponding Cutoff SHMFs because transfer function suppression begins at larger scales in WDM. This is consistent with the comparisons between DM models with different transfer function shapes and suppression scales in \cite{Nadler241003635}.

\end{document}